\documentclass[11pt]{article}

\usepackage{amsmath}
\usepackage{amssymb}
\usepackage{graphicx}
\usepackage{subfigure}

\def\middlespace {\smallskipamount=5.625pt plus1.5pt minus1.5pt
                  \medskipamount=11.25pt plus3pt minus3pt
                  \bigskipamount=22.5pt plus6pt minus6pt
                  \normalbaselineskip=22.5pt plus0pt minus0pt
                  \normallineskip=1pt
                  \normallineskiplimit=0pt
                  \jot=5.625pt
                  {\def\smallskip {\vskip\smallskipamount}}
                  {\def\medskip   {\vskip\medskipamount}}
                  {\def\bigskip   {\vskip\bigskipamount}}
                  {\setbox\strutbox=\hbox{\vrule
                    height15.75pt depth6.75pt width 0pt}}
                  \parskip 11.25pt
                  \normalbaselines}

\begin{document}

\ \vskip 1.0 in

\begin{center}
 { \Large {\bf Einstein Gravity as the Thermodynamic Limit }}

\smallskip

{\Large {\bf  of an Underlying Quantum Statistics}}

\vskip 0.2 in

\smallskip

\bigskip

\bigskip

\bigskip

{{\large
{\bf T. P. Singh\footnote{e-mail address: tpsingh@tifr.res.in} 
} 
}}

\medskip

{\it Tata Institute of Fundamental Research,}\\
{\it Homi Bhabha Road, Mumbai 400 005, India.}\\
\medskip

\vskip 0.5cm
\end{center}

\vskip 1.0 in

\begin{abstract}

\noindent The black hole area theorem suggests that classical general relativity is the thermodynamic
limit of a quantum statistics. The degrees of freedom of the statistical theory cannot be the spacetime
metric. We argue that the statistical theory should be constructed from a noncommutative gravity, whose
classical, and thermodynamic, approximation is Einstein gravity. The noncommutative gravity theory
exhibits a duality between quantum fields and macroscopic black holes, which is used to show that the
black hole possesses an entropy of the order of its area. The principle on which this work is based
also provides a possible explanation for the smallness of the cosmological constant, and for the quantum
measurement problem, indicating that this is a promising avenue towards the merger of quantum mechanics
and gravity.

\vskip 1.0 in

\end{abstract}

\newpage

\middlespace

\noindent 

\noindent
The black hole area theorem says that in any process involving the
interaction of black holes, the sum of the areas of their horizons can 
never decrease. This is puzzling. How can it be that classical general relativity, which
is supposedly a theory invariant under time reversal, possesses such an irreversible feature? The only plausible answer seems to be that the gravitational
degrees of freedom, although generally regarded as dynamical, are in reality
only thermodynamic entities. They are the macroscopic manifestation of
an underlying microscopic theory whose degrees of freedom are not the
spacetime metric. Similar ideas have been suggested and developed by various 
authors in the past \cite{various}. In the present note we propose a
quantum statistics to which general relativity could be a thermodynamic
approximation, and we use it to explain why the 
black hole has an entropy, which is proportional to its area. 

Additional support for a thermodynamic interpretation of gravity comes
of course from Hawking radiation, and from the proportionality of the 
Bekenstein-Hawking entropy to the area of the horizon. Thus
\begin{equation}
\rm{Entropy}\ \propto \rm{Area}\ \propto\ \ln \Omega(q),
\label{dof}
\end{equation}
where $\Omega(q)$ is the number of microstates for which the thermodynamic
variables take their most probable value, in equilibrium. 
In accordance with the conventional interpretation of thermodynamics, 
the quantum mechanical
degrees of freedom $q$, naturally cannot include the spacetime metric \footnote 
{Contrast this with the entropy of the electromagnetic blackbody radiation - the
degrees of freedom $q$ now are indeed the electromagnetic field, but the entropy
is not expressible in terms of the classical electrodynamic field.}. 
Neither can these degrees of freedom be guessed at from a knowledge of Einstein
gravity (one cannot derive statistical mechanics starting from thermodynamics). 
The starting point for the underlying quantum statistical mechanics must be found
elsewhere. 

Over the last few years we have emphasized that there should exist a reformulation
of quantum theory which does not make reference to an external classical spacetime \cite{singh}.
An attractive candidate for the mathematical development of such a reformulation is
noncommutative differential geometry. One introduces  a set of noncommuting coordinates
$(\hat{x},\hat{t})$ which satisfy appropriate commutation relations, and a corresponding
noncommutative line-element with an antisymmetric part
\begin{equation}
ds^{2}= \hat{g}_{00}d\hat{t}^{2}-\hat{g}_{11}d\hat{x}^{2} + 
\hat{\theta} (d\hat{t}d\hat{x}-d\hat{x}d\hat{t}).
\end{equation}
The resulting noncommutative gravity theory is assumed to be invariant under
general coordinate transformations of the noncommuting coordinates. In the limit
that the mass-energy acting as the source becomes much smaller than Planck mass,
all the above metric components reduce to unity. Dynamics in this `quantum Minkowski spacetime'
is described by a noncommutative special relativity, and can be shown to reduce to
standard quantum mechanics as and when an external spacetime becomes available. 

In the opposite limit, when the mass-energy acting as the source is much greater than 
Planck mass, $\hat{\theta}$ goes to zero, the coordinates become commuting, the line-element becomes
symmetric, and the noncommutative gravity theory reduces to general relativity. In the intermediate
regime, where the mass-energy is comparable to Planck mass, the antisymmetric component $\hat{\theta}$
makes the theory different from Einstein gravity, introducing crucial degrees of freedom not present when the metric is symmetric. It is worth emphasizing that the antisymmetric part of the metric is forced on us by the need for
a reformulation of quantum mechanics. 

We are now ready to see how a thermodynamic interpretation for Einstein gravity arises in an inevitable
and elegant manner on macroscopic scales, from the underlying noncommutative theory. For a set of microstates labeled by the noncommuting metric $\hat{h}_{\mu\nu}\equiv(\hat{g}_{\mu\nu},\hat{\theta})$, where the components of $\hat{h}_{\mu\nu}$ do not commute with each other, the thermodynamic entropy can be written as    
\begin{equation}
S = \ln \Omega (\hat{h}_{\mu\nu}).
\label{sgrav}
\end{equation}
By construction, in the macroscopic limit, the thermodynamic entropy makes no reference to the underlying
microscopic degrees of freedom $\hat{h}_{\mu\nu}$. On the other hand, it is central to our analysis that
the degrees $\hat{h}_{\mu\nu}$, though fundamentally different from the classical metric $g_{\mu\nu}$, 
reduce to the latter in the macroscopic, classical approximation $\hat{\theta}\rightarrow 0$. 
The macroscopic system is described by $g_{\mu\nu}$, and by the entropy $S$.
One could hence either say that $g_{\mu\nu}$ are thermodynamical variables, which obey a thermodynamics constructed from the above entropy function; or one could say that the $g_{\mu\nu}$ obey Einstein's field equations. It is
impossible to distinguish the two viewpoints. The interpretation of gravity as spacetime curvature is a remnant of
the general covariance of the underlying noncommutative theory. 

It is crucial to understand why the thermodynamic interpretation is unique to gravity, and will
not hold for any other interaction, say electrodynamics. Let us express the entropy of the
electromagnetic field $F_{\mu\nu}$ in terms of the quantum microstates as
\begin{equation}
S = \ln \Omega (F_{\mu\nu}).
\end{equation}  
Now, in the macroscopic limit, the thermodynamic entropy makes no reference to the underlying
microscopic degrees of freedom $F_{\mu\nu}$. Moreover, it is this same field
which appears at both the microscopic and macroscopic levels. Thus, a thermodynamic picture
for electrodynamics is not possible. In contrast, for gravity, the fundamental variables $\hat{h}_{\mu\nu}$
differ from the coarse grained variables $g_{\mu\nu}$, resulting in a thermodynamic viewpoint. 
The thermodynamics determined for the metric variables $g_{\mu\nu}$ by the entropy function (\ref{sgrav})
is equivalent to Einstein gravity precisely because general relativity is the classical limit of the
noncommutative gravity theory on which (\ref{sgrav}) is based. Notably, the system behaves like a non-degenerate
gas, because the thermodynamic approximation is the same as the classical approximation.

On the basis of these ideas we can now demonstrate why a semiclassical black hole,
with a mass much greater than Planck mass $m_{Pl}$, has an entropy proportional
to its area. In recent work \cite{singh2} based on the above mentioned noncommutative gravity theory, we have argued that there exists a duality between the states of a macroscopic black hole of mass $M$, and those of a corresponding quantum field whose particles have a mass $m=m_{Pl}^{2}/M$. The energy levels of the quantum field extend up to
Planck mass, in steps of $m$ [as opposed to being a continuum], so that the number of levels  is 
$m_{Pl}/m=M/m_{Pl}$. The energy levels for the dual black hole extend further, up to M, and hence the
number of such levels is
\begin{equation} 
N=\frac{M}{m}=\frac{M^{2}}{m_{Pl}^{2}}.
\label{enn}
\end{equation}
To a very good approximation, this then is the number of `eigenstates' of the black hole $M$,
by virtue of the duality between the `weakly quantum, strongly gravitational' black hole, and the
`strongly quantum, weakly gravitational' quantum field. These eigenstates are labeled by the noncommutative
metric $\hat{h}_{\mu\nu}$ and {\it not} by the classical metric $g_{\mu\nu}$ - which is precisely the reason for the
emergent thermodynamic nature of gravity.

The mass $M$ of the black hole is made up of $N=M/m=M^{2}/m_{Pl}^{2}$ fundamental indistinguishable mass quanta $m$,
each of which can go into any of the $N$ distinguishable eigenstates. From the viewpoint of a microcanonical ensemble,
this can be done in ${}^{(2N-1)}C_{N}$ different ways,
which for large $N$ approximates to $\exp[2N\ln 2]\equiv \Omega(q)$. The entropy of the black hole is therefore given by
\begin{equation}
S=\ln \Omega(q) =2\ln 2\ \frac{M^{2}}{m_{Pl}^2}\sim 0.03 \times {\rm Area}
\end{equation}
which establishes the proportionality of the entropy to the area. [We can hardly expect to recover the celebrated
coefficient of $1/4$ from this approximate calculation]! The temperature of the black hole, statistically defined
as $(\partial S/\partial M)^{-1}$, comes out inversely proportional to $M$, as expected.

We can also deduce the relation between entropy and area by noting that the black hole, being macroscopic
and nearly classical, should obey Maxwell-Boltzmann statistics. If the quanta $m$ are treated as distinguishable
then each of the $N$ quanta can go into any of the $N$ states. This gives rise to $N^{N}$ different combinations.
We must divide this number by the factor $N!$ to avoid the Gibbs paradox, so that the resulting entropy is an
extensive quantity (increasing the number of quanta $m$ by a factor $f$ should increase the entropy by the same factor).  
Hence  the number of microstates for the black hole is
\begin{equation}
\Omega(q) = \frac{{N}^{N}}{N!}= \frac{1} {(M^{2}/m_{Pl}^2)!}
\left(\frac{M^2}{m_{Pl}^2}\right)^{M^{2}/m_{Pl}^{2}}\sim \exp{[M^{2}/m_{Pl}^{2}]} \quad 
{\rm for}\ M\gg m_{Pl}.
\label{fun}
\end{equation}
This again leads to the conclusion that the entropy $S=\ln \Omega(q)$ is proportional to the
area $M^{2}$ of the black hole. Duality plays a central role in arriving at this inference. 
(The number $\Omega(q)$ in (\ref{fun}) has the interesting form $p^{p}/p!$, which goes to $e^{p}$ for
large $p$). We also observe that the entropy is indeed extensive, being proportional to $N$.
Retaining a higher order term in the Stirling approximation for the denominator in (\ref{fun})
gives a (negative) logarithmic correction to the Bekenstein-Hawking entropy formula :
\begin{equation}
S \sim \frac{M^{2}}{m_{Pl}^{2}} - \ln \frac{M^{2}}{m_{Pl}^{2}}.
\end{equation}

The quantum-classical duality also makes it apparent why the black hole entropy is extensive, in spite
of being proportional to the area, rather than the volume. Ordinary, non-gravitational, entropy is extensive 
and proportional to the volume of the system because there is no duality principle operating there.

In view of the above results, and the advocated thermodynamic interpretation
of general relativity, the area theorem is to be viewed as a thermodynamic
remnant of the underlying quantum statistics. It is of course apparent that
the `mechanical' part of the thermodynamics comes from the remaining
gravitational degrees of freedom, such as gravitational waves. It is also
evident that our explicit construction of the entropy applies only for the
equilibrium case. Einstein gravity in general is equivalent to a 
non-equilibrium thermodynamics. 

The principle that there should be a reformulation of quantum mechanics which does not
refer to an external classical time has far-reaching implications. It strongly suggests
the generalization of Einstein gravity to a noncommutative gravity, and a consequent
duality between black holes and quantum fields. As we have seen here,
it provides a statistical underpinning for the thermodynamic interpretation of general
relativity. The aforementioned principle also
suggests a nonlinear generalization of quantum mechanics which could explain the collapse of the
wave-function during a quantum measurement - an experimentally testable prediction \cite{singh3}.
It is remarkable that a construction very similar to the one used in this essay
to calculate entropy also explains why the Universe should have a non-zero but tiny value 
for the cosmological constant, consistent with astronomical observations \cite{singh2}.
Perhaps, in the broad synthesis of ideas and predictions that is emerging in this program, we might finally
have a testable glimpse of the unification of quantum mechanics with gravity.

\end{document}